\documentclass[graphicx]{revtex4}
\usepackage{graphicx}
\usepackage{color}
\begin{document}

\title{Efficient entanglement purification for polarization logic Bell state with the photonic Faraday rotation}
\author{Lan Zhou$^{1,2}$ and Yu-Bo Sheng$^{2\ast}$}

\address{$^1$College of Mathematics \& Physics, Nanjing University of Posts and Telecommunications, Nanjing,
210003, China\\
$^2$Key Lab of Broadband Wireless Communication and Sensor Network
 Technology,
 Nanjing University of Posts and Telecommunications, Ministry of
 Education, Nanjing, 210003,
 China\\
\footnote{Email address: shengyb@njupt.edu.cn}}

\begin{abstract}
Logic-qubit entanglement is a promising resource in quantum information processing, especially in future large-scale quantum networks. In the paper,  we put forward an efficient entanglement purification protocol (EPP) for nonlocal mixed logic entangled states with the bit-flip error in the logic qubits of the logic Bell state, resorting to the photon-atom interaction in low-quality (Q) cavity and atomic state measurement. Different from  existing EPPs, this protocol can also purify the logic phase-flip error, and the bit-flip error and the phase-flip error in physic qubit.  During the protocol, we only require to measure the atom states, and it is useful for improving the entanglement of photon systems in future large-scale quantum networks.
\end{abstract}
 \maketitle

\section{Introduction}
Entanglement is an indispensable resource which is widely applied in many aspects, such as quantum teleportation \cite{teleportation}, quantum key distribution (QKD)\cite{Ekert91}, quantum secret sharing (QSS)\cite{QSS}, quantum secure direct communication (QSDC) \cite{QSDC1,QSDC2}.  During the past two decades,  many types of entanglement were investigated for quantum communication, such as  polarization entanglement \cite{ghz}, time-bin entanglement, hybrid entanglement, hyperentanglement, and so on.
In 2011, Fro\"{w}is and D\"{u}r investigated a new type of entanglement \cite{cghz1}. It is the logic-qubit entanglement, which encodes many physic qubits in a logic qubit. It is also called concatenated Greenberger-Horne-Zeilinger (C-GHZ) state.  The typical C-GHZ state can be written as \cite{cghz1,cghz2,cghz3,cghz4,yan,pan1,logicbell1,logicbell2,logicconcentration}
\begin{eqnarray}
|\Phi^{\pm}\rangle_{N,M}=\frac{1}{\sqrt{2}}(|GHZ^{+}_{M}\rangle^{\otimes N} \pm |GHZ^{-}_{M}\rangle^{\otimes N}),\label{logic}
\end{eqnarray}
where $N$ and $M$ are the number of logic qubits and the number of physic qubits in each logic qubit, respectively. In Eq. (\ref{logic}), the logic qubit $|GHZ^{+}_{M}\rangle$ is a GHZ state of the form
\begin{eqnarray}
|GHZ^{\pm}_{M}\rangle=\frac{1}{\sqrt{2}}(|0\rangle^{\otimes M}\pm |1\rangle^{\otimes M}).
\end{eqnarray}
In 2014,  Lu \emph{et al.} first experimentally generated the logic-qubit entangled state with $M=2$ and $N=3$ in linear optics \cite{pan1}.  They also showed that the C-GHZ state is useful for large-scale fibre-based quantum networks and
multipartite QKD, QSS and third-man quantum cryptography. There are some other important progresses for C-GHZ state. For example, in 2013, Ding \emph{et al.} described an interesting approach to prepare the C-GHZ state, resorting to the cross-Kerr nonlinearity \cite{yan}.  In 2015, Sheng \emph{et al.} firstly proposed the logic Bell-state analysis  and arbitrary C-GHZ state analysis protocols with the help of the controlled-not (CNOT) gate \cite{logicbell1} and nonlinear optical elements \cite{logicbell2}, respectively.
Based on the logic Bell-state analysis and arbitrary C-GHZ state analysis, they also described the teleportation of a logic qubit and the approach of logic entanglement swapping. Their protocols show that it is possible to set up the long-distance quantum channel based on the logic-qubit entanglement.

 Unfortunately, the entanglement  is generally fragile, where noise and decoherence can diminish or even destroy the desirable quantum features. In the applications, a degraded quantum channel may make the fidelity of the teleportation degrade, even more, it will make the quantum communication  insecure.
 The logic-qubit entangled state may also suffer from the decoherence. It will make the maximally logic-qubit entangled state degrade to the mixed state. Prior to the application, we have to recover the degraded entangled states into the maximally entangled states. As one of the key techniques, the entanglement purification can distill the high quality entangled states from the low quality entangled states \cite{purification1,addpurification2,purification2,purification3,purification4,purification5,purification6,purification7,purification8,purification9,purification10,
purification11,purification12,purification13,purification14,purification15,addpurification,purification16,purification17,purification18,purification19,purification20,purification21,purification22,EPP1,EPP2,EPP3}. In 1996, Bennett \emph{et al.} put forward the first entanglement purification protocol (EPP) for the Werner state with the help of the controlled-not (CNOT) gate.\cite{purification1}. In the same year, this protocol was improved by Deustch \emph{et al.} with similar
quantum logical operations \cite{addpurification2}. Subsequently, many efficient EPPs have been proposed. For example, in 2001, the group of Pan described
 a feasible EPP with linear optics \cite{purification3}. The entanglement purification for multi-particle and high dimension systems were also proposed \cite{EPP1,EPP2,EPP3}. In 2008, Sheng \emph{et al.} described a recyclable EPP which can obtain a higher fidelity \cite{purification6}.  In 2010, they proposed the deterministic EPPs \cite{purification7,purification8}.  In 2014, the efficient EPP for hyperentanglement ware also proposed \cite{purification15}. Recently, the EPPs for the noisy blind quantum computation were also presented \cite{addpurification,purification17}. On the other hand, there are also some efficient EPPs for the solid quantum systems, such as the EPP for spins \cite{purification18}, short chains of atoms \cite{purification21,purification22}.

Existing EPPs  cannot deal with the logic-qubit entanglement. Compared with the conventional physic Bell state,  logic Bell state has more complex structure. Moreover, the mixed state of logic Bell state contains more errors than the mixed state of conventional physic Bell state.
The mixed state of logic Bell state contains not only the bit-flip error and phase-flip error in the logic qubits,  but also the bit-flip error and the phase-flip error in physic qubit. In this paper, we put forward an efficient EPP for the polarization logic Bell state,  resorting to the photonic Faraday rotation.  We will show that both the bit-flip error and phase-flip error in logic-qubit entanglement can be well purified. On the other hand, if a bit-flip error occurs
in one physic qubit, we can completely correct it. The phase-flip error in one physic qubit equals to the bit-flip error in the logic-qubit entanglement, which can also be well purified.

This paper is organized as follows: In Sec. 2, we briefly introduce the basic principle of the photonic Faraday rotation. In Sec. 3, we explain
the purification for the bit-flip error and phase-flip error in the logic-qubit entanglement. In Sec. 4,
we describe the purification for the physic-qubit error. In
Sec. 5, we present a discussion and conclusion.

\section{Basic principle of the photonic Faraday rotation}
The quantum electrodynamics (QED) is a promising
platform for performing the quantum information tasks due to
the controllable interaction between atoms and photons. For a long time, with the atoms strongly interacting
with local high-quality (Q) cavities, the spatially separated cavities could serve as quantum
nodes, and construct a quantum network assisted by the photons acting as a quantum bus \cite{QED1,QED2,QED3,QED4}. However, the requirements for high-Q cavities and strong coupling to the confined atoms are stringent for current techniques. Fortunately, in 2009, the group of An successfully implemented the quantum information processing (QIP) tasks with the moderate cavity-atom coupling
in the low-Q cavities \cite{cavity}. This method works in the low-Q cavities and only involves
the virtual excitation of atoms. Therefore, it is insensitive to both
the cavity decay and the atomic spontaneous emission. Following this scheme, various works based on the QED in low-Q cavity
have been presented \cite{gate,gate1,cavity1,zhouatom,zhouentropy,wei1,wei2,zhoupra2}.

\begin{figure}[!h]
\begin{center}
\includegraphics[width=8cm,angle=0]{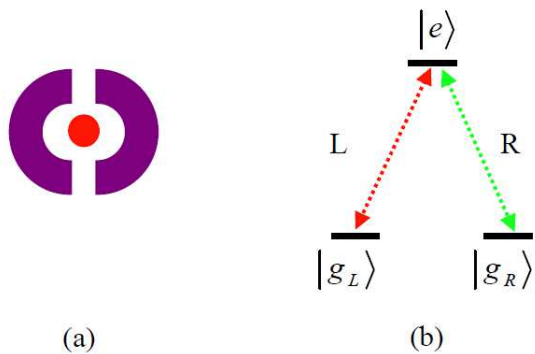}
\caption{The basic principle of the interaction between the photon pulse and the three-level atom in the
low-Q cavity \cite{zhoupra2}. a): A three-level atom is trapped in a low-Q cavity. b): The three-level atom has an excited state $|e\rangle$ and two degenerate ground states $|g_{L}\rangle$ and $|g_{R}\rangle$, respectively. The state $|g_{L}\rangle$ and $|g_{R}\rangle$ couple with a left (L) polarized and a right
(R) polarized photon, respectively .}
\end{center}
\end{figure}
The atomic structure and the interaction between the photon pulse and the three-level atom in the
low-Q cavity are shown in Fig. 1 \cite{faraday,faraday1,faraday2,faraday3}. We make a three-level atom trap in the low-Q cavity. The atom has two degenerate ground states $|g_{L}\rangle$ and $|g_{R}\rangle$
and an excited state $|e\rangle$. The transition between $|g_{L}\rangle$ and $|e\rangle$ is assisted with a left-circularly polarized
photon ($|L\rangle$), while that between $|g_{R}\rangle$ and $|e\rangle$ is assisted with a right-circularly polarized
photon ($|R\rangle$), respectively. A single photon pulse with frequency $\omega_{p}$ enters the optical cavity. Using the adiabatic approximation,  we can solve the Langevin equations of
motion for cavity and atomic lowering operators analytically. Then, we obtain the general expression of the reflection coefficient of the atom-cavity
system in the form of \cite{cavity1,rotation1,rotation2}
\begin{eqnarray}
r(\omega_{p})\equiv\frac{a_{out}(t)}{a_{in}(t)}=\frac{[i(\omega_{c}-\omega_{p})-\frac{\kappa}{2}][i(\omega_{0}-\omega_{p})+\frac{\gamma}{2}]
+\lambda^{2}}{[i(\omega_{c}-\omega_{p})+\frac{\kappa}{2}][i(\omega_{0}-\omega_{p})+\frac{\gamma}{2}]+\lambda^{2}}.\label{reflect1}
\end{eqnarray}
 Here, $a_{in}(t)$ and $a_{out}(t)$ are the cavity input operator and cavity output operator, respectively. $\kappa$ and $\gamma$ are the cavity damping rate and atomic decay rate. $\omega_{p}$, $\omega_{c}$, and $\omega_{0}$ are the frequency of the input photon, the cavity, and the atom, respectively. $\lambda$ is the atom-cavity coupling strength.

 If the atom uncouples to the cavity, which makes $\lambda=0$, Eq. (\ref{reflect1}) for an empty cavity can be simplified as
\begin{eqnarray}
r_{0}(\omega_{p})=\frac{i(\omega_{c}-\omega_{p})-\frac{\kappa}{2}}{i(\omega_{c}-\omega_{p})+\frac{\kappa}{2}}.\label{r0}
\end{eqnarray}
$r_{0}(\omega_{p})$ can be written as as a pure phase shift as $r_{0}(\omega_{p})=e^{i\theta_{0}}$. On the other hand, for $r(\omega_{p})$, as the photon experiences an extremely weak absorption in the interaction process, we consider that the output photon only experiences a pure phase shift without any absorption as a good approximation. Consequently, with strong $\kappa$ and weak $\gamma$ and $\lambda$, we can rewrite Eq. (\ref{reflect1}) as $r(\omega_{p})\simeq e^{i\theta}$.

Especially, under a special condition that $\omega_{0}=\omega_{c}$, $\omega_{p}=\omega_{c}-\frac{\kappa}{2}$, and $\lambda=\frac{\kappa}{2}$, we can
obtain $\theta=\pi$ and $\theta_{0}=\frac{\pi}{2}$. Therefore, when the photon is reflected from the low-Q cavity, we can obtain the relationship between the input and output photon combined with the atomic state as \cite{zhouatom,zhouentropy,Pengpra}
\begin{eqnarray}
|L\rangle|g_{L}\rangle\rightarrow-|L\rangle|g_{L}\rangle, \quad |R\rangle|g_{L}\rangle\rightarrow i|R\rangle|g_{L}\rangle,\nonumber\\
|L\rangle|g_{R}\rangle\rightarrow i|L\rangle|g_{R}\rangle, \quad |R\rangle|g_{R}\rangle\rightarrow -|R\rangle|g_{R}\rangle.\label{relation}
\end{eqnarray}

\section{The purification of the logic-entanglement}
\subsection{The purification for the bit-flip error}

We first introduce the purification for the logic Bell state under the simplest case, that is, $M=2$. We suppose two parties, say Alice and Bob share a maximally entangled logic Bell state $|\Phi^{+}\rangle_{AB}$ with the form of
\begin{eqnarray}
|\Phi^{\pm}\rangle_{AB}=\frac{1}{\sqrt{2}}(|\phi^{+}\rangle_{A}|\phi^{+}\rangle_{B}\pm|\phi^{-}\rangle_{A}|\phi^{-}\rangle_{B}).\label{initial}
\end{eqnarray}
 Here the subscripts $A$ and $B$ means Alice and Bob, respectively. If a logic bit-flip error occurs  with the probability of $1-F$, it will change $|\Phi^{+}\rangle_{AB}$ to $|\Psi^{+}\rangle_{AB}$ as
\begin{eqnarray}
|\Psi^{\pm}\rangle_{AB}=\frac{1}{\sqrt{2}}(|\phi^{+}\rangle_{A}|\phi^{-}\rangle_{B}\pm|\phi^{-}\rangle_{A}|\phi^{+}\rangle_{B}).\label{biterror}
\end{eqnarray}
In Eq. (\ref{initial}) and Eq. (\ref{biterror}), $|\phi^{\pm}\rangle$ are two of the four polarization Bell states. The four polarization Bell states can be written as
\begin{eqnarray}
|\phi^{\pm}\rangle=\frac{1}{\sqrt{2}}(|L\rangle|L\rangle\pm|R\rangle|R\rangle),\nonumber\\
|\psi^{\pm}\rangle=\frac{1}{\sqrt{2}}(|L\rangle|R\rangle\pm|R\rangle|L\rangle).
\end{eqnarray}
Due to the bit-flip error, the initial photon state degrades to a mixed state as
\begin{eqnarray}
\rho_{in1}=F|\Phi^{+}\rangle_{AB}\langle\Phi^{+}|+(1-F)|\Psi^{+}\rangle_{AB}\langle\Psi^{+}|.\label{mix}
\end{eqnarray}

\begin{figure}[!h]
\begin{center}
\includegraphics[width=10cm,angle=0]{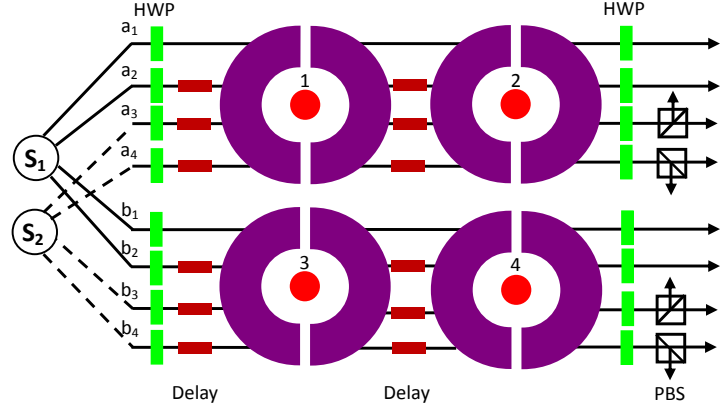}
\caption{The schematic drawing of  EPP for logic bit-flip error of the logic Bell state with $M=2$. Two  copies of the mixed photon states are generated from the photon sources S1 and S2, respectively. The parties need to prepare four three-level atoms with the form of $\frac{1}{\sqrt{2}}(|g_{L}\rangle+|g_{R}\rangle)$ trapped in four low-Q cavities, respectively. HWP represents the half-wave plate, and BS means the 50:50 beam splitter. The "Delay" setup is used to ensure each cavity only contains a photon at a time.}
\end{center}
\end{figure}

For realizing the entanglement purification, Alice and Bob require to share two pairs of the mixed states in Eq. (\ref{mix}), here named $\rho_{in1}$ and $\rho_{in2}$. $\rho_{in1}$ is in the spatial modes of $a_{1}$, $a_{2}$, $b_{1}$, and $b_{2}$, while $\rho_{in2}$ is in the spatial modes $a_{3}$, $a_{4}$, $b_{3}$, and $b_{4}$. The whole state of $\rho_{in1}\otimes\rho_{in2}$ can be described as follows. The whole state is in the state $|\Phi^{+}\rangle_{A1B1}\otimes|\Phi^{+}\rangle_{A2B2}$ with the probability of $F^{2}$. It is in the state  $|\Phi^{+}\rangle_{A1B1}\otimes|\Psi^{+}\rangle_{A2B2}$ or $|\Psi^{+}\rangle_{A1B1}\otimes|\Phi^{+}\rangle_{A2B2}$ with the equal probability of $F(1-F)$. With the probability of $(1-F)^{2}$, it is in the state $|\Psi^{+}\rangle_{A1B1}\otimes|\Psi^{+}\rangle_{A2B2}$.

As shown in Fig. 2, before purification, the parties make each of the photons pass through a half-wave plate (HWP), which will make $|L\rangle\rightarrow\frac{1}{\sqrt{2}}(|L\rangle+|R\rangle)$ and $|R\rangle\rightarrow\frac{1}{\sqrt{2}}(|L\rangle-|R\rangle)$.  After the HWP, $|\phi^{+}\rangle$ will not change, but $|\phi^{-}\rangle$ will change to $|\psi^{+}\rangle$. Therefore, $|\Phi^{+}\rangle_{AB}$ and $|\Psi^{+}\rangle_{AB}$ will evolve to
\begin{eqnarray}
|\Phi^{+}\rangle_{AB}&\rightarrow&|\Phi'^{+}\rangle_{AB}=\frac{1}{\sqrt{2}}(|\phi^{+}\rangle_{A}|\phi^{+}\rangle_{B}+|\psi^{+}\rangle_{A}|\psi^{+}\rangle_{B}),\\
|\Psi^{+}\rangle_{AB}&\rightarrow&|\Psi'^{+}\rangle_{AB}=\frac{1}{\sqrt{2}}(|\phi^{+}\rangle_{A}|\psi^{+}\rangle_{B}+|\psi^{+}\rangle_{A}|\phi^{+}\rangle_{B}).
\end{eqnarray}

 Then, the parties prepare four three-level atoms in the same states of $|\Omega^{\pm}_{i}\rangle=\frac{1}{\sqrt{2}}(|g_{L}\rangle\pm|g_{R}\rangle)$ $(i=1, 2, 3, 4)$. The four atoms here named atom $"1"$, $"2"$, $"3"$, and $"4"$ are trapped in four low-Q cavities, respectively. They make the photons in the $a_{1}a_{2}$ and $a_{3}a_{4}$ modes pass through two low-Q cavities and successively interact with atom "1" and "2", the photons in the $b_{1}b_{2}$ and $b_{3}b_{4}$ modes successively enter two cavities and interact with atom "3" and "4", respectively. It is noticed that the parties should ensure that each cavity only contains one photon at a time, which should be well controlled
in practical experiment. In our protocol, the "Delay" setup is adopted to complete this task. For example, they first make the
photon in the $a_{1}$ mode enter the cavity and interact with the atom "1". After the photon is reflected and exits the cavity, they let the
photon in the $a_{2}$ mode enter the cavity. Based on the photon-atom interaction rules in Eq. (\ref{relation}), the parties can obtain
\begin{eqnarray}
&&|\phi^{+}\rangle|\phi^{+}\rangle\otimes\frac{1}{\sqrt{2}}(|g_{L}\rangle+|g_{R}\rangle)\otimes\frac{1}{\sqrt{2}}(|g_{L}\rangle+|g_{R}\rangle)
\rightarrow|\phi^{+}\rangle|\phi^{+}\rangle\otimes\frac{1}{\sqrt{2}}(|g_{L}\rangle+|g_{R}\rangle)\otimes\frac{1}{\sqrt{2}}(|g_{L}\rangle+|g_{R}\rangle),\\
&&|\phi^{+}\rangle|\psi^{+}\rangle\otimes\frac{1}{\sqrt{2}}(|g_{L}\rangle+|g_{R}\rangle)\otimes\frac{1}{\sqrt{2}}(|g_{L}\rangle+|g_{R}\rangle)
\rightarrow-|\phi^{+}\rangle|\psi^{+}\rangle\otimes\frac{1}{\sqrt{2}}(|g_{L}\rangle-|g_{R}\rangle)\otimes\frac{1}{\sqrt{2}}(|g_{L}\rangle-|g_{R}\rangle),\\
&&|\psi^{+}\rangle|\phi^{+}\rangle\otimes\frac{1}{\sqrt{2}}(|g_{L}\rangle+|g_{R}\rangle)\otimes\frac{1}{\sqrt{2}}(|g_{L}\rangle+|g_{R}\rangle)
\rightarrow-|\psi^{+}\rangle|\phi^{+}\rangle\otimes\frac{1}{\sqrt{2}}(|g_{L}\rangle-|g_{R}\rangle)\otimes\frac{1}{\sqrt{2}}(|g_{L}\rangle-|g_{R}\rangle),\\
&&|\psi^{+}\rangle|\psi^{+}\rangle\otimes\frac{1}{\sqrt{2}}(|g_{L}\rangle+|g_{R}\rangle)\otimes\frac{1}{\sqrt{2}}(|g_{L}\rangle+|g_{R}\rangle)
\rightarrow|\psi^{+}\rangle|\psi^{+}\rangle\otimes\frac{1}{\sqrt{2}}(|g_{L}\rangle+|g_{R}\rangle)\otimes\frac{1}{\sqrt{2}}(|g_{L}\rangle+|g_{R}\rangle).
\end{eqnarray}

In this way, after all the photons are reflected from the two cavities, if the initial state is $|\Phi'^{+}\rangle_{A1B1}\otimes|\Phi'^{+}\rangle_{A2B2}$ with the probability of $F^{2}$, the parties can obtain
\begin{eqnarray}
&&|\Phi'^{+}\rangle_{A1B1}\otimes|\Phi'^{+}\rangle_{A2B2}\otimes|\Omega^{+}_{1}\rangle\otimes|\Omega^{+}_{2}\rangle\otimes|\Omega^{+}_{3}\rangle\otimes|\Omega^{+}_{4}\rangle\\
&=&\frac{1}{\sqrt{2}}(|\phi^{+}\rangle_{a1a2}|\phi^{+}\rangle_{b1b2}+|\psi^{+}\rangle_{a1a2}|\psi^{+}\rangle_{b1b2})\otimes
\frac{1}{\sqrt{2}}(|\phi^{+}\rangle_{a3a4}|\phi^{+}\rangle_{b3b4}+|\psi^{+}\rangle_{a3a4}|\psi^{+}\rangle_{b3b4})
\otimes|\Omega^{+}_{1}\rangle\otimes|\Omega^{+}_{2}\rangle\otimes|\Omega^{+}_{3}\rangle\otimes|\Omega^{+}_{4}\rangle\nonumber\\
&\rightarrow&\frac{1}{2}[|\phi^{+}\rangle_{a1a2}|\phi^{+}\rangle_{a3a4}|\phi^{+}\rangle_{b1b2}|\phi^{+}\rangle_{b3b4}
|\Omega^{+}_{1}\rangle|\Omega^{+}_{2}\rangle|\Omega^{+}_{3}\rangle|\Omega^{+}_{4}\rangle
+|\phi^{+}\rangle_{a1a2}|\psi^{+}\rangle_{a3a4}|\phi^{+}\rangle_{b1b2}|\psi^{+}\rangle_{b3b4}
|\Omega^{-}_{1}\rangle|\Omega^{-}_{2}\rangle|\Omega^{-}_{3}\rangle|\Omega^{-}_{4}\rangle\nonumber\\
&+&|\psi^{+}\rangle_{a1a2}|\phi^{+}\rangle_{a3a4}|\psi^{+}\rangle_{b1b2}|\phi^{+}\rangle_{b3b4}
|\Omega^{-}_{1}\rangle|\Omega^{-}_{2}\rangle|\Omega^{-}_{3}\rangle|\Omega^{-}_{4}\rangle
+|\psi^{+}\rangle_{a1a2}|\psi^{+}\rangle_{a3a4}|\psi^{+}\rangle_{b1b2}|\psi^{+}\rangle_{b3b4}
|\Omega^{+}_{1}\rangle|\Omega^{+}_{2}\rangle|\Omega^{+}_{3}\rangle|\Omega^{+}_{4}\rangle].\label{1}
\end{eqnarray}

If the initial state is $|\Phi'^{+}\rangle_{A1B1}\otimes|\Psi'^{+}\rangle_{A2B2}$ or $|\Psi'^{+}\rangle_{A1B1}\otimes|\Phi'^{+}\rangle_{A2B2}$ with the equal probability of $F(1-F)$, they will obtain
\begin{eqnarray}
&&|\Phi'^{+}\rangle_{A1B1}\otimes|\Psi'^{+}\rangle_{A2B2}\otimes
|\Omega^{+}_{1}\rangle\otimes|\Omega^{+}_{2}\rangle\otimes|\Omega^{+}_{3}\rangle\otimes|\Omega^{+}_{4}\rangle\nonumber\\
&=&\frac{1}{\sqrt{2}}(|\phi^{+}\rangle_{a1a2}|\phi^{+}\rangle_{b1b2}+|\psi^{+}\rangle_{a1a2}|\psi^{+}\rangle_{b1b2})\otimes
\frac{1}{\sqrt{2}}(|\phi^{+}\rangle_{a3a4}|\psi^{+}\rangle_{b3b4}+|\psi^{+}\rangle_{a3a4}|\phi^{+}\rangle_{b3b4})
\otimes|\Omega^{+}_{1}\rangle\otimes|\Omega^{+}_{2}\rangle\otimes|\Omega^{+}_{3}\rangle\otimes|\Omega^{+}_{4}\rangle\nonumber\\
&\rightarrow&\frac{1}{2}[|\phi^{+}\rangle_{a1a2}|\phi^{+}\rangle_{a3a4}|\phi^{+}\rangle_{b1b2}|\psi^{+}\rangle_{b3b4}
|\Omega^{+}_{1}\rangle|\Omega^{+}_{2}\rangle|\Omega^{-}_{3}\rangle|\Omega^{-}_{4}\rangle
+|\phi^{+}\rangle_{a1a2}|\psi^{+}\rangle_{a3a4}|\phi^{+}\rangle_{b1b2}|\phi^{+}\rangle_{b3b4}
|\Omega^{-}_{1}\rangle|\Omega^{-}_{2}\rangle|\Omega^{+}_{3}\rangle|\Omega^{+}_{4}\rangle\nonumber\\
&+&|\psi^{+}\rangle_{a1a2}|\phi^{+}\rangle_{a3a4}|\psi^{+}\rangle_{b1b2}|\psi^{+}\rangle_{b3b4}
|\Omega^{-}_{1}\rangle|\Omega^{-}_{2}\rangle|\Omega^{+}_{3}\rangle|\Omega^{+}_{4}\rangle
+|\psi^{+}\rangle_{a1a2}|\psi^{+}\rangle_{a3a4}|\psi^{+}\rangle_{b1b2}|\phi^{+}\rangle_{b3b4}
|\Omega^{+}_{1}\rangle|\Omega^{+}_{2}\rangle|\Omega^{-}_{3}\rangle|\Omega^{-}_{4}\rangle],\label{2}
\end{eqnarray}
and
\begin{eqnarray}
&&|\Psi'^{+}\rangle_{A1B1}\otimes|\Phi'^{+}\rangle_{A2B2}\otimes|\Omega^{+}_{1}\rangle\otimes|\Omega^{+}_{2}\rangle\otimes
|\Omega^{+}_{3}\rangle\otimes|\Omega^{+}_{4}\rangle\nonumber\\
&=&\frac{1}{\sqrt{2}}(|\phi^{+}\rangle_{a1a2}|\psi^{+}\rangle_{b1b2}+|\psi^{+}\rangle_{a1a2}|\phi^{+}\rangle_{b1b2})\otimes
\frac{1}{\sqrt{2}}(|\phi^{+}\rangle_{a3a4}|\phi^{+}\rangle_{b3b4}+|\psi^{+}\rangle_{a3a4}|\psi^{+}\rangle_{b3b4})
\otimes|\Omega^{+}_{1}\rangle\otimes|\Omega^{+}_{2}\rangle\otimes|\Omega^{+}_{3}\rangle\otimes|\Omega^{+}_{4}\rangle\nonumber\\
&\rightarrow&\frac{1}{2}[|\phi^{+}\rangle_{a1a2}|\phi^{+}\rangle_{a3a4}|\psi^{+}\rangle_{b1b2}|\phi^{+}\rangle_{b3b4}
|\Omega^{+}_{1}\rangle|\Omega^{+}_{2}\rangle|\Omega^{-}_{3}\rangle|\Omega^{-}_{4}\rangle
+|\phi^{+}\rangle_{a1a2}|\psi^{+}\rangle_{a3a4}|\psi^{+}\rangle_{b1b2}|\psi^{+}\rangle_{b3b4}
|\Omega^{-}_{1}\rangle|\Omega^{-}_{2}\rangle
|\Omega^{+}_{3}\rangle|\Omega^{+}_{4}\rangle\nonumber\\
&+&|\psi^{+}\rangle_{a1a2}|\phi^{+}\rangle_{a3a4}|\phi^{+}\rangle_{b1b2}|\phi^{+}\rangle_{b3b4}|\Omega^{-}_{1}\rangle|\Omega^{-}_{2}\rangle
|\Omega^{+}_{3}\rangle|\Omega^{+}_{4}\rangle
+|\psi^{+}\rangle_{a1a2}|\psi^{+}\rangle_{a3a4}|\phi^{+}\rangle_{b1b2}|\psi^{+}\rangle_{b3b4}|\Omega^{+}_{1}\rangle|\Omega^{+}_{2}\rangle
|\Omega^{-}_{3}\rangle|\Omega^{-}_{4}\rangle].\label{3}
\end{eqnarray}

  For the initial state of $|\Psi'^{+}\rangle_{A1B1}\otimes|\Psi'^{+}\rangle_{A2B2}$ with the probability of $(1-F)^{2}$, the whole state will evolve to
\begin{eqnarray}
&&|\Psi'^{+}\rangle_{A1B1}\otimes|\Psi'^{+}\rangle_{A2B2}\otimes|\Omega^{+}_{1}\rangle\otimes|\Omega^{+}_{2}\rangle\otimes|\Omega^{+}_{3}\rangle\otimes|\Omega^{+}_{4}\rangle\nonumber\\
&=&\frac{1}{\sqrt{2}}(|\phi^{+}\rangle_{a1a2}|\psi^{+}\rangle_{b1b2}+|\psi^{+}\rangle_{a1a2}|\phi^{+}\rangle_{b1b2})\otimes
\frac{1}{\sqrt{2}}(|\phi^{+}\rangle_{a3a4}|\psi^{+}\rangle_{b3b4}+|\psi^{+}\rangle_{a3a4}|\phi^{+}\rangle_{b3b4})
\otimes|\Omega^{+}_{1}\rangle\otimes|\Omega^{+}_{2}\rangle\otimes|\Omega^{+}_{3}\rangle\otimes|\Omega^{+}_{4}\rangle\nonumber\\
&\rightarrow&\frac{1}{2}[|\phi^{+}\rangle_{a1a2}|\phi^{+}\rangle_{a3a4}|\psi^{+}\rangle_{b1b2}|\psi^{+}\rangle_{b3b4}
|\Omega^{+}_{1}\rangle|\Omega^{+}_{2}\rangle|\Omega^{+}_{3}\rangle|\Omega^{+}_{4}\rangle
+|\phi^{+}\rangle_{a1a2}|\psi^{+}\rangle_{a3a4}|\psi^{+}\rangle_{b1b2}|\phi^{+}\rangle_{b3b4}
|\Omega^{-}_{1}\rangle|\Omega^{-}_{2}\rangle|\Omega^{-}_{3}\rangle|\Omega^{-}_{4}\rangle\nonumber\\
&+&|\psi^{+}\rangle_{a1a2}|\phi^{+}\rangle_{a3a4}|\phi^{+}\rangle_{b1b2}|\psi^{+}\rangle_{b3b4}
|\Omega^{-}_{1}\rangle|\Omega^{-}_{2}\rangle|\Omega^{-}_{3}\rangle|\Omega^{-}_{4}\rangle
+|\psi^{+}\rangle_{a1a2}|\psi^{+}\rangle_{a3a4}|\phi^{+}\rangle_{b1b2}|\phi^{+}\rangle_{b3b4}
|\Omega^{+}_{1}\rangle|\Omega^{+}_{2}\rangle|\Omega^{+}_{3}\rangle|\Omega^{+}_{4}\rangle].\label{4}
\end{eqnarray}

After all the photons exiting the cavities, the parties perform the
Hadamard operation on the four atoms, which makes $|g_{L}\rangle\rightarrow\frac{1}{\sqrt{2}}(|g_{L}\rangle+|g_{R}\rangle)$, and  $|g_{R}\rangle\rightarrow\frac{1}{\sqrt{2}}(|g_{L}\rangle-|g_{R}\rangle)$. After that, it can be found that $|\Omega^{+}\rangle\rightarrow|g_{L}\rangle$ and $|\Omega^{-}\rangle\rightarrow|g_{R}\rangle$. Next, the parties measure the states of the four atoms in the basis of $\{|g_{L}\rangle, |g_{R}\rangle\}$. From Eq. (\ref{1}) to Eq. (\ref{4}), if the measurement results of the four atoms are the same, say, $|g_{L}\rangle_{1}|g_{L}\rangle_{2}|g_{L}\rangle_{3}|g_{L}\rangle_{4}$ or $|g_{R}\rangle_{1}|g_{R}\rangle_{2}|g_{R}\rangle_{3}|g_{R}\rangle_{4}$, our purification protocol is successful, while if the measurement results of atoms "1" and "2" are different with the atoms "3" and "4", say $|g_{L}\rangle_{1}|g_{L}\rangle_{2}|g_{R}\rangle_{3}|g_{R}\rangle_{4}$ or $|g_{R}\rangle_{1}|g_{R}\rangle_{2}|g_{L}\rangle_{3}|g_{L}\rangle_{4}$, the purification protocol fails.

For example, suppose that the measurement results are $|g_{L}\rangle_{1}|g_{L}\rangle_{2}|g_{L}\rangle_{3}|g_{L}\rangle_{4}$. They will obtain
the state
\begin{eqnarray}
\frac{1}{\sqrt{2}}(|\phi^{+}\rangle_{a1a2}|\phi^{+}\rangle_{b1b2}|\phi^{+}\rangle_{a3a4}|\phi^{+}\rangle_{b3b4}
+|\psi^{+}\rangle_{a1a2}|\psi^{+}\rangle_{b1b2}|\psi^{+}\rangle_{a3a4}|\psi^{+}\rangle_{b3b4}),\label{add1}
\end{eqnarray}
with the probability of $F^{2}$, and obtain the state
\begin{eqnarray}
\frac{1}{\sqrt{2}}(|\phi^{+}\rangle_{a1a2}|\psi^{+}\rangle_{b1b2}|\phi^{+}\rangle_{a3a4}|\psi^{+}\rangle_{b3b4}
+|\psi^{+}\rangle_{a1a2}|\phi^{+}\rangle_{b1b2}|\psi^{+}\rangle_{a3a4}|\phi^{+}\rangle_{b3b4}),\label{add2}
\end{eqnarray}
with the probability of $(1-F)^{2}$.
In order to obtain the output mixed state with the same form of Eq. (\ref{mix}), the parties first make all the photons in spatial modepass through the HWPs, which can transmit $|\psi^{+}\rangle$ to $|\phi^{-}\rangle$, and keep $|\phi^{+}\rangle$ constant. Therefore, the state in Eq. (\ref{add1}) becomes
\begin{eqnarray}
&&\frac{1}{\sqrt{2}}(|\phi^{+}\rangle_{a1a2}|\phi^{+}\rangle_{b1b2}|\phi^{+}\rangle_{a3a4}|\phi^{+}\rangle_{b3b4}
+|\phi^{-}\rangle_{a1a2}|\phi^{-}\rangle_{b1b2}|\phi^{-}\rangle_{a3a4}|\phi^{-}\rangle_{b3b4})\nonumber\\
&=&\frac{1}{\sqrt{2}}[|\phi^{+}\rangle_{a1a2}|\phi^{+}\rangle_{b1b2}\otimes\frac{1}{2}(|LLLL\rangle_{a3a4b3b4}+|LLRR\rangle_{a3a4b3b4}+|RRLL\rangle_{a3a4b3b4}+|RRRR\rangle_{a3a4b3b4})\nonumber\\
&+&|\phi^{-}\rangle_{a1a2}|\phi^{-}\rangle_{b1b2}\otimes\frac{1}{2}(|LLLL\rangle_{a3a4b3b4}-|LLRR\rangle_{a3a4b3b4}-|RRLL\rangle_{a3a4b3b4}+|RRRR\rangle_{a3a4b3b4})].\label{add3}
\end{eqnarray}
The state in Eq. (\ref{add2}) becomes
\begin{eqnarray}
&&\frac{1}{\sqrt{2}}(|\phi^{+}\rangle_{a1a2}|\phi^{-}\rangle_{b1b2}|\phi^{+}\rangle_{a3a4}|\phi^{-}\rangle_{b3b4}
+|\phi^{-}\rangle_{a1a2}|\phi^{+}\rangle_{b1b2}|\phi^{-}\rangle_{a3a4}|\phi^{+}\rangle_{b3b4})\nonumber\\
&=&\frac{1}{\sqrt{2}}[|\phi^{+}\rangle_{a1a2}|\phi^{-}\rangle_{b1b2}\otimes\frac{1}{2}(|LLLL\rangle_{a3a4b3b4}-|LLRR\rangle_{a3a4b3b4}+|RRLL\rangle_{a3a4b3b4}-|RRRR\rangle_{a3a4b3b4})\nonumber\\
&+&|\phi^{-}\rangle_{a1a2}|\phi^{+}\rangle_{b1b2}\otimes\frac{1}{2}(|LLLL\rangle_{a3a4b3b4}+|LLRR\rangle_{a3a4b3b4}-|RRLL\rangle_{a3a4b3b4}-|RRRR\rangle_{a3a4b3b4})]\label{add4}
\end{eqnarray}
Subsequently, they let the four photons in $a3$, $a4$, $b3$ and $b4$ pass through the polarization beam splitters (PBSs), which can transmit the $|L\rangle$ photon and reflect the $|R\rangle$ photon, respectively. Finally, they measure the four photons. If the measurement are the same, they are
$|LLLL\rangle$ or $|RRRR\rangle$, they will finally obtain a new mixed state
\begin{eqnarray}
\rho_{out1}=F'|\Phi^{+}\rangle_{A1B1}\langle\Phi^{+}|+(1-F')|\Psi^{+}\rangle_{A1B1}\langle\Psi^{+}|.
\end{eqnarray}
Here $F'=\frac{F^{2}}{F^{2}+(1-F)^{2}}$.
It is obvious that $\rho_{out1}$ has the same form of $\rho_{in1}$. It can be calculated that $F'>F$ under $F>\frac{1}{2}$.
on the other hand, if the measurement results are different, they are $|LLRR\rangle$ or $|RRLL\rangle$, they will finally obtain another new mixed state
\begin{eqnarray}
\rho'_{out1}=F'|\Phi^{-}\rangle_{A1B1}\langle\Phi^{-}|+(1-F')|\Psi^{-}\rangle_{A1B1}\langle\Psi^{-}|.
\end{eqnarray}
State $\rho'_{out1}$ can be transformed to $\rho_{out1}$ by performing the bit-flip operations on all the physical qubits in one of the logic qubit.
 So far, the purification has been successfully completed.

Interestingly, this purification protocol can be extended to the logic Bell states with each logic qubit being  arbitrary GHZ state. Suppose Alice and Bob share the state as
\begin{eqnarray}
|\Phi^{+}_{M}\rangle_{AB}=\frac{1}{\sqrt{2}}(|GHZ^{+}_{M}\rangle_{A}|GHZ^{+}_{M}\rangle_{B}+|GHZ^{-}_{M}\rangle_{A}|GHZ^{-}_{M}\rangle_{B}),\label{initialN}
\end{eqnarray}
where
\begin{eqnarray}
|GHZ^{\pm}_{M}\rangle=\frac{1}{\sqrt{2}}(|L\rangle^{\otimes M}\pm|R\rangle^{\otimes M}).
\end{eqnarray}
If the bit-flip error occurs with the probability of $(1-F)$, $|\Phi^{+}_{M}\rangle_{AB}$ will convert to $|\Psi^{+}_{M}\rangle_{AB}$ with the form of
\begin{eqnarray}
|\Psi^{+}_{M}\rangle_{AB}=\frac{1}{\sqrt{2}}(|GHZ^{+}_{M}\rangle_{A}|GHZ^{-}_{M}\rangle_{B}+|GHZ^{-}_{M}\rangle_{A}|GHZ^{+}_{M}\rangle_{B}).\label{initialN}
\end{eqnarray}
In this way, Alice and Bob share a mixed state as
\begin{eqnarray}
\rho_{M}=F|\Phi^{+}_{M}\rangle_{AB}\langle\Phi^{+}_{M}|+(1-F)|\Psi^{+}_{M}\rangle_{AB}\langle\Psi^{+}_{M}|.\label{mixN}
\end{eqnarray}

\begin{figure}[!h]
\begin{center}
\includegraphics[width=10cm,angle=0]{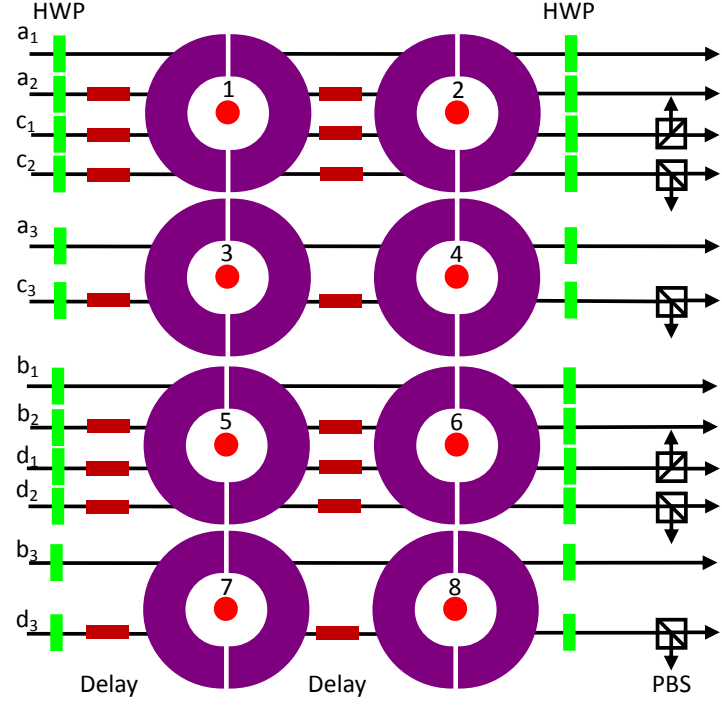}
\caption{The schematic drawing of the EPP for logic bit-flip error of the logic Bell state with $M=3$. Two same copies of the mixed photon states are required. The parties need to prepare eight three-level atoms with the form of $\frac{1}{\sqrt{2}}(|g_{L}\rangle+|g_{R}\rangle)$ trapped in eight low-Q cavities, respectively. }
\end{center}
\end{figure}

\begin{figure}[!h]
\begin{center}
\includegraphics[width=10cm,angle=0]{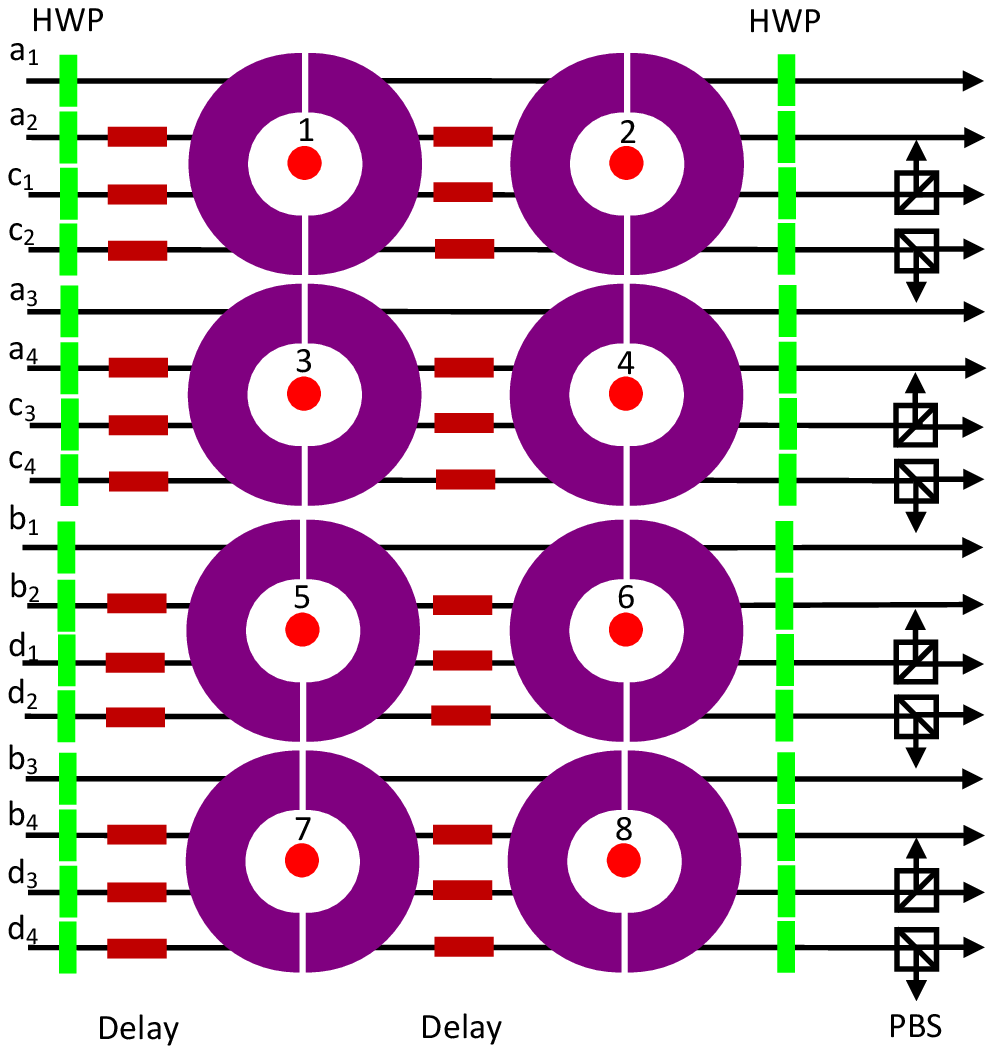}
\caption{The schematic drawing of the EPP for logic bit-flip error of the logic Bell state with $M=4$. The parties also need to prepare eight three-level atoms with the form of $\frac{1}{\sqrt{2}}(|g_{L}\rangle+|g_{R}\rangle)$ trapped in eight low-Q cavities, respectively. }
\end{center}
\end{figure}
For completing the purification task, Alice and Bob also require to share two same copies of the mixed states as shown in Eq. (\ref{mixN}). The first copy of mixed state is in the spatial modes $a_{1}$, $b_{1}$, $a_{2}$, $b_{2}$, $\cdots$, $a_{m}$, $b_{m}$, and the second copy of mixed state is in the spatial modes $c_{1}$, $d_{1}$, $c_{2}$, $d_{2}$, $\cdots$, $c_{m}$, $d_{m}$, respectively. Then, the whole photon state can be written as follows. It is in the state of $|\Phi^{+}_{M}\rangle_{AB}\otimes|\Phi^{+}_{M}\rangle_{CD}$ with the probability of $F^{2}$. It is in the state of $|\Phi^{+}_{M}\rangle_{AB}\otimes|\Psi^{+}_{M}\rangle_{CD}$ or $|\Psi^{+}_{M}\rangle_{AB}\otimes|\Phi^{+}_{M}\rangle_{CD}$ with the equal probability of $F(1-F)$. With the probability of $(1-F)^{2}$, it is in the state of $|\Psi^{+}_{M}\rangle_{AB}\otimes|\Psi^{+}_{M}\rangle_{CD}$. For realizing the purification, the parties need to first perform the Hadamard operations on all the photons. We take the Hadamard operations on $|GHZ^{\pm_{M}}\rangle_{A}$ for example. Suppose the parties first make the photons in $a_{1}a_{2}$ modes pass through the HWPs. After that, they can obtain
\begin{eqnarray}
|GHZ^{+}_{M}\rangle_{A}
&\rightarrow&\frac{1}{\sqrt{2}}[\frac{1}{\sqrt{2}}(|L\rangle_{a1}+|R\rangle_{a1})\frac{1}{\sqrt{2}}(|L\rangle_{a2}+|R\rangle_{a2})|L\rangle^{\otimes M-2}_{a3\cdots am}+\frac{1}{\sqrt{2}}(|L\rangle_{a1}-|R\rangle_{a1})\frac{1}{\sqrt{2}}(|L\rangle_{a2}-|R\rangle_{a2})|R\rangle^{\otimes M-2}_{a3\cdots am}]\nonumber\\
&=&\frac{1}{\sqrt{2}}(|\phi^{+}\rangle_{a1a2}|GHZ^{+}_{M-2}\rangle_{a3\cdots am}+|\psi^{+}\rangle_{a1a2}|GHZ^{-}_{M-2}\rangle_{a3\cdots am}),\\
|GHZ^{-}_{M}\rangle_{A}
&\rightarrow&\frac{1}{\sqrt{2}}[\frac{1}{\sqrt{2}}(|L\rangle_{a1}+|R\rangle_{a1})\frac{1}{\sqrt{2}}(|L\rangle_{a2}+|R\rangle_{a2})|L\rangle^{\otimes M-2}_{a3\cdots am}-\frac{1}{\sqrt{2}}(|L\rangle_{a1}-|R\rangle_{a1})\frac{1}{\sqrt{2}}(|L\rangle_{a2}-|R\rangle_{a2})|R\rangle^{\otimes M-2}_{a3\cdots am}]\nonumber\\
&=&\frac{1}{\sqrt{2}}(|\phi^{+}\rangle_{a1a2}|GHZ^{-}_{M-2}\rangle_{a3\cdots am}+|\psi^{+}\rangle_{a1a2}|GHZ^{+}_{M-2}\rangle_{a3\cdots am}).
\end{eqnarray}
Then, the parties make the photons in the $a_{3}a_{4}$ modes pass through the HWPs, which will make
\begin{eqnarray}
|GHZ^{+}_{M}\rangle_{A}
&\rightarrow&\frac{1}{\sqrt{2}}[|\phi^{+}\rangle_{a1a2}\frac{1}{2}(|\phi^{+}\rangle_{a3a4}|GHZ^{+}_{M-4}\rangle_{a5\cdots am}+|\psi^{+}\rangle_{a3a4}|GHZ^{-}_{M-4}\rangle_{a5\cdots am})\nonumber\\
&+&|\psi^{+}\rangle_{a1a2}\frac{1}{2}(|\phi^{+}\rangle_{a3a4}|GHZ^{-}_{M-4}\rangle_{a5\cdots am}+|\psi^{+}\rangle_{a3a4}|GHZ^{+}_{M-4}\rangle_{a5\cdots am})]\nonumber\\
&=&\frac{1}{2}[(|\phi^{+}\rangle_{a1a2}|\phi^{+}\rangle_{a3a4}+|\psi^{+}\rangle_{a1a2}|\psi^{+}\rangle_{a3a4})|GHZ^{+}_{M-4}\rangle_{a5\cdots am}\nonumber\\
&+&(|\phi^{+}\rangle_{a1a2}|\psi^{+}\rangle_{a3a4}+|\psi^{+}\rangle_{a1a2}|\phi^{+}\rangle_{a3a4})|GHZ^{-}_{M-4}\rangle_{a5\cdots am}],\\
|GHZ^{-}_{M}\rangle_{A}
&\rightarrow&\frac{1}{_{2}}[|\phi^{+}\rangle_{a1a2}\frac{1}{2}(|\phi^{+}\rangle_{a3a4}|GHZ^{-}_{M-4}\rangle_{a5\cdots am}+|\psi^{+}\rangle_{a3a4}|GHZ^{+}_{M-4}\rangle_{a5\cdots am})\nonumber\\
&+&|\psi^{+}\rangle_{a1a2}\frac{1}{2}(|\phi^{+}\rangle_{a3a4}|GHZ^{-}_{M-4}\rangle_{a5\cdots am}+|\psi^{+}\rangle_{a3a4}|GHZ^{+}_{M-4}\rangle_{a5\cdots am})]\nonumber\\
&=&\frac{1}{2}[(|\phi^{+}\rangle_{a1a2}|\phi^{+}\rangle_{a3a4}+|\psi^{+}\rangle_{a1a2}|\psi^{+}\rangle_{a3a4})|GHZ^{-}_{M-4}\rangle_{a5\cdots am}\nonumber\\
&+&(|\phi^{+}\rangle_{a1a2}|\psi^{+}\rangle_{a3a4}+|\psi^{+}\rangle_{a1a2}|\phi^{+}\rangle_{a3a4})|GHZ^{+}_{M-4}\rangle_{a5\cdots am}].
\end{eqnarray}
Similarly, after we make the remaining $m-4$ photons pass through the HWPs two by two, we can obtain the factorization of $|GHZ^{\pm}_{M}\rangle_{A}$. For when $M$ is odd or even, factorization of $|GHZ^{\pm}_{M}\rangle$ is slightly different. When $M$ is odd, the last items of the iteration are $|GHZ^{\pm}_{1}\rangle_{a_{m}}$. After the HWPs, we can obtain
\begin{eqnarray}
|GHZ^{+}_{1}\rangle_{A}\rightarrow|L\rangle_{a_{m}}, \quad
|GHZ^{-}_{1}\rangle_{A}\rightarrow|R\rangle_{a_{m}}.
\end{eqnarray}
When $M$ is even, the last items of the iteration are $|GHZ^{\pm}_{2}\rangle_{a_{m-1}a_{m}}$. After the HWPs, they can evolve to
\begin{eqnarray}
|GHZ^{+}_{2}\rangle_{A}\rightarrow|\phi^{+}\rangle_{a_{m-1}a_{m}}, \quad
|GHZ^{-}_{2}\rangle_{A}\rightarrow|\psi^{+}\rangle_{a_{m-1}a_{m}}.
\end{eqnarray}

For explaining the purification process in detail, we take the cases with $M=3$ and $M=4$ for example. The purification processes can be straightly extend to the cases with arbitrary odd $M$ and even $M$, respectively.
 Under $M=3$, after the HWPs, we can obtain
\begin{eqnarray}
|GHZ^{+}_{3}\rangle\rightarrow\frac{1}{\sqrt{2}}(|\phi^{+}\rangle|L\rangle+|\psi^{+}\rangle|R\rangle),\quad
|GHZ^{-}_{3}\rangle\rightarrow\frac{1}{\sqrt{2}}(|\phi^{+}\rangle|R\rangle+|\psi^{+}\rangle|L\rangle).
\end{eqnarray}

As shown in Fig. 3, the parties need to prepare eight three-level atoms with the form of $\frac{1}{\sqrt{2}}(|g_{L}\rangle+|g_{R}\rangle)$ trapped in eight low-Q cavities, respectively. They make the photons in the $a_{1}a_{2}c_{1}c_{2}$ and $b_{1}b_{2}d_{1}d_{2}$ modes pass through cavities and successively interact with atom "1" "2", and "5" "6", respectively, the photons in the $a_{3}c_{3}$  and $b_{3}d_{3}$ modes pass through cavities and successively interact with the atom "3" "4" and "7" "8", respectively. According to the input-output relationship in Eq. (\ref{relation}), we can also obtain
\begin{eqnarray}
&&|LL\rangle\otimes\frac{1}{\sqrt{2}}(|g_{L}\rangle+|g_{R}\rangle)\otimes\frac{1}{\sqrt{2}}(|g_{L}\rangle+|g_{R}\rangle)
\rightarrow|LL\rangle\otimes\frac{1}{\sqrt{2}}(|g_{L}\rangle-|g_{R}\rangle)\otimes\frac{1}{\sqrt{2}}(|g_{L}\rangle-|g_{R}\rangle),\nonumber\\
&&|RR\rangle\otimes\frac{1}{\sqrt{2}}(|g_{L}\rangle+|g_{R}\rangle)\otimes\frac{1}{\sqrt{2}}(|g_{L}\rangle+|g_{R}\rangle)
\rightarrow|RR\rangle\otimes\frac{1}{\sqrt{2}}(|g_{L}\rangle-|g_{R}\rangle)\otimes\frac{1}{\sqrt{2}}(|g_{L}\rangle-|g_{R}\rangle),\nonumber\\
&&|LR\rangle\otimes\frac{1}{\sqrt{2}}(|g_{L}\rangle+|g_{R}\rangle)\otimes\frac{1}{\sqrt{2}}(|g_{L}\rangle+|g_{R}\rangle)
\rightarrow -|LR\rangle\otimes\frac{1}{\sqrt{2}}(|g_{L}\rangle+|g_{R}\rangle)\otimes\frac{1}{\sqrt{2}}(|g_{L}\rangle+|g_{R}\rangle),\nonumber\\
&&|RL\rangle\otimes\frac{1}{\sqrt{2}}(|g_{L}\rangle+|g_{R}\rangle)\otimes\frac{1}{\sqrt{2}}(|g_{L}\rangle+|g_{R}\rangle)
\rightarrow -|RL\rangle\otimes\frac{1}{\sqrt{2}}(|g_{L}\rangle+|g_{R}\rangle)\otimes\frac{1}{\sqrt{2}}(|g_{L}\rangle+|g_{R}\rangle).\label{relation1}
\end{eqnarray}
 After all the photons exiting the cavity, they make a Hadamard operation on all the eight atoms and measure them in the basis of $\{|g_{L}\rangle, |g_{R}\rangle\}$. If the measurement results of the eight atoms are one of the eight cases, say $|g_{L}\rangle_{1}|g_{L}\rangle_{2}|g_{R}\rangle_{3}|g_{R}\rangle_{4}|g_{L}\rangle_{5}|g_{L}\rangle_{6}|g_{R}\rangle_{7}|g_{R}\rangle_{8}$, $|g_{L}\rangle_{1}|g_{L}\rangle_{2}|g_{R}\rangle_{3}|g_{R}\rangle_{4}|g_{R}\rangle_{5}|g_{R}\rangle_{6}|g_{L}\rangle_{7}|g_{L}\rangle_{8}$,
$|g_{R}\rangle_{1}|g_{R}\rangle_{2}|g_{L}\rangle_{3}|g_{L}\rangle_{4}|g_{L}\rangle_{5}|g_{L}\rangle_{6}|g_{R}\rangle_{7}|g_{R}\rangle_{8}$, $|g_{R}\rangle_{1}|g_{R}\rangle_{2}|g_{L}\rangle_{3}|g_{L}\rangle_{4}|g_{R}\rangle_{5}|g_{R}\rangle_{6}|g_{L}\rangle_{7}|g_{L}\rangle_{8}$,\\
$|g_{L}\rangle_{1}|g_{L}\rangle_{2}|g_{L}\rangle_{3}|g_{L}\rangle_{4}|g_{L}\rangle_{5}|g_{L}\rangle_{6}|g_{L}\rangle_{7}|g_{L}\rangle_{8}$,
$|g_{L}\rangle_{1}|g_{L}\rangle_{2}|g_{L}\rangle_{3}|g_{L}\rangle_{4}|g_{R}\rangle_{5}|g_{R}\rangle_{6}|g_{R}\rangle_{7}|g_{R}\rangle_{8}$,\\
$|g_{R}\rangle_{1}|g_{R}\rangle_{2}|g_{R}\rangle_{3}|g_{R}\rangle_{4}|g_{L}\rangle_{5}|g_{L}\rangle_{6}|g_{L}\rangle_{7}|g_{L}\rangle_{8}$, and
$|g_{R}\rangle_{1}|g_{R}\rangle_{2}|g_{R}\rangle_{3}|g_{R}\rangle_{4}|g_{R}\rangle_{5}|g_{R}\rangle_{6}|g_{R}\rangle_{7}|g_{R}\rangle_{8}$, the purification will be successful. On the other hand, if the measurement results of the eight atoms are one of the eight cases,
$|g_{L}\rangle_{1}|g_{L}\rangle_{2}|g_{R}\rangle_{3}|g_{R}\rangle_{4}|g_{L}\rangle_{5}|g_{L}\rangle_{6}|g_{L}\rangle_{7}|g_{L}\rangle_{8}$, $|g_{L}\rangle_{1}|g_{L}\rangle_{2}|g_{R}\rangle_{3}|g_{R}\rangle_{4}|g_{R}\rangle_{5}|g_{R}\rangle_{6}|g_{R}\rangle_{7}|g_{R}\rangle_{8}$,
$|g_{R}\rangle_{1}|g_{R}\rangle_{2}|g_{L}\rangle_{3}|g_{L}\rangle_{4}|g_{L}\rangle_{5}|g_{L}\rangle_{6}|g_{L}\rangle_{7}|g_{L}\rangle_{8}$, $|g_{R}\rangle_{1}|g_{R}\rangle_{2}|g_{L}\rangle_{3}|g_{L}\rangle_{4}|g_{R}\rangle_{5}|g_{R}\rangle_{6}|g_{R}\rangle_{7}|g_{R}\rangle_{8}$,\\
$|g_{L}\rangle_{1}|g_{L}\rangle_{2}|g_{L}\rangle_{3}|g_{L}\rangle_{4}|g_{L}\rangle_{5}|g_{L}\rangle_{6}|g_{R}\rangle_{7}|g_{R}\rangle_{8}$,
$|g_{L}\rangle_{1}|g_{L}\rangle_{2}|g_{L}\rangle_{3}|g_{L}\rangle_{4}|g_{R}\rangle_{5}|g_{R}\rangle_{6}|g_{L}\rangle_{7}|g_{L}\rangle_{8}$,\\
$|g_{R}\rangle_{1}|g_{R}\rangle_{2}|g_{R}\rangle_{3}|g_{R}\rangle_{4}|g_{L}\rangle_{5}|g_{L}\rangle_{6}|g_{R}\rangle_{7}|g_{R}\rangle_{8}$, and
$|g_{R}\rangle_{1}|g_{R}\rangle_{2}|g_{R}\rangle_{3}|g_{R}\rangle_{4}|g_{R}\rangle_{5}|g_{R}\rangle_{6}|g_{L}\rangle_{7}|g_{L}\rangle_{8}$,
the purification will fail.

We take the successful case $|g_{L}\rangle_{1}|g_{L}\rangle_{2}|g_{R}\rangle_{3}|g_{R}\rangle_{4}|g_{L}\rangle_{5}|g_{L}\rangle_{6}|g_{R}\rangle_{7}|g_{R}\rangle_{8}$ for example. Under this case, the parties will obtain
\begin{eqnarray}
&&\frac{1}{2}[(|\phi^{+}\rangle_{a1a2}|L\rangle_{a3}|\phi^{+}\rangle_{c1c2}|L\rangle_{c3}+|\psi^{+}\rangle_{a1a2}|R\rangle_{a3}|\psi^{+}\rangle_{c1c2}|R\rangle_{c3})
(|\phi^{+}\rangle_{b1b2}|L\rangle_{b3}|\phi^{+}\rangle_{d1d2}|L\rangle_{d3}+|\psi^{+}\rangle_{b1b2}|R\rangle_{b3}|\psi^{+}\rangle_{d1d2}|R\rangle_{d3})\nonumber\\
&+&(|\phi^{+}\rangle_{a1a2}|R\rangle_{a3}|\phi^{+}\rangle_{c1c2}|R\rangle_{c3}+|\psi^{+}\rangle_{a1a2}|L\rangle_{a3}|\psi^{+}\rangle_{c1c2}|L\rangle_{c3})
(|\phi^{+}\rangle_{b1b2}|R\rangle_{b3}|\phi^{+}\rangle_{d1d2}|R\rangle_{d3}+|\psi^{+}\rangle_{b1b2}|L\rangle_{b3}|\psi^{+}\rangle_{d1d2}|L\rangle_{d3})],\nonumber\\
\end{eqnarray}
with the probability of $F^{2}$, or obtain
\begin{eqnarray}
&&\frac{1}{2}[(|\phi^{+}\rangle_{a1a2}|L\rangle_{a3}|\phi^{+}\rangle_{c1c2}|L\rangle_{c3}+|\psi^{+}\rangle_{a1a2}|R\rangle_{a3}|\psi^{+}\rangle_{c1c2}|R\rangle_{c3})
(|\phi^{+}\rangle_{b1b2}|R\rangle_{b3}|\phi^{+}\rangle_{d1d2}|R\rangle_{d3}+|\psi^{+}\rangle_{b1b2}|L\rangle_{b3}|\psi^{+}\rangle_{d1d2}|L\rangle_{d3})\nonumber\\
&+&(|\phi^{+}\rangle_{a1a2}|R\rangle_{a3}|\phi^{+}\rangle_{c1c2}|R\rangle_{c3}+|\psi^{+}\rangle_{a1a2}|L\rangle_{a3}|\psi^{+}\rangle_{c1c2}|L\rangle_{c3})
(|\phi^{+}\rangle_{b1b2}|L\rangle_{b3}|\phi^{+}\rangle_{d1d2}|L\rangle_{d3}+|\psi^{+}\rangle_{b1b2}|R\rangle_{b3}|\psi^{+}\rangle_{d1d2}|R\rangle_{d3})],\nonumber\\
\end{eqnarray}
with the probability of $(1-F)^{2}$.

Next, the parties make all the photons pass through the HWPs. Then, they make each of the photons in the $c_{1}c_{2}c_{3}$ and $d_{1}d_{2}d_{3}$ modes pass through a PBS, and detect the photons in the output modes. After the detection, the parties can finally obtain 
\begin{eqnarray}
\rho_{out3}=F'|\Phi^{+}_{3}\rangle_{AB}\langle\Phi^{+}_{3}|+(1-F')|\Psi^{+}_{3}\rangle_{AB}\langle\Psi^{+}_{3}|
\end{eqnarray}
or 
\begin{eqnarray}
\rho_{out4}=F'|\Phi^{-}_{3}\rangle_{AB}\langle\Phi^{-}_{3}|+(1-F')|\Psi^{-}_{3}\rangle_{AB}\langle\Psi^{-}_{3}|
\end{eqnarray}
with $F'=\frac{F^{2}}{F^{2}+(1-F)^{2}}$.
If they obtain $\rho_{out4}$, they can transform it to $\rho_{out3}$ by performing the bit-flip operations on all the physical qubits in one of the logic qubit. Similarly, if the parties get other seven successful cases, they will obtain the same mixed states as $\rho_{out3}$ and $\rho_{out4}$. So far, the purification process for the mixed state with $M=3$ is completed.

Under $M=4$, after passing through the HWPs, $|GHZ^{\pm}_{4}\rangle$ will evolve to
\begin{eqnarray}
|GHZ^{+}_{4}\rangle\rightarrow\frac{1}{\sqrt{2}}(|\phi^{+}\rangle|\phi^{+}\rangle+|\psi^{+}\rangle|\psi^{+}\rangle),\quad
|GHZ^{-}_{4}\rangle\rightarrow\frac{1}{\sqrt{2}}(|\phi^{+}\rangle|\psi^{+}\rangle+|\psi^{+}\rangle|\phi^{+}\rangle).
\end{eqnarray}

As shown in Fig. 4, the parties also prepare eight three-level atoms with $\frac{1}{\sqrt{2}}(|g_{L}\rangle+|g_{R}\rangle)$ in eight low-Q cavities, respectively. The photons can be divided into four groups, that is, the photons in $a_{1}a_{2}c_{1}c_{2}$, $a_{3}a_{4}c_{3}c_{4}$, $b_{1}b_{2}d_{1}d_{2}$ and $b_{3}b_{4}d_{3}d_{4}$. The parties make the photons in each group pass through two cavities and interact with two atoms, successively. After all the photons exiting the cavities, they make a Hadamard operation on the eight atoms and measure them in the basis of $\{|g_{L}\rangle, |g_{R}\rangle\}$. Similar with the case of $M=3$, if the parties obtain the measurement results of $|g_{L}\rangle_{1}|g_{L}\rangle_{2}|g_{R}\rangle_{3}|g_{R}\rangle_{4}|g_{L}\rangle_{5}|g_{L}\rangle_{6}|g_{R}\rangle_{7}|g_{R}\rangle_{8}$, $|g_{L}\rangle_{1}|g_{L}\rangle_{2}|g_{R}\rangle_{3}|g_{R}\rangle_{4}|g_{R}\rangle_{5}|g_{R}\rangle_{6}|g_{L}\rangle_{7}|g_{L}\rangle_{8}$,
$|g_{R}\rangle_{1}|g_{R}\rangle_{2}|g_{L}\rangle_{3}|g_{L}\rangle_{4}|g_{L}\rangle_{5}|g_{L}\rangle_{6}|g_{R}\rangle_{7}|g_{R}\rangle_{8}$, $|g_{R}\rangle_{1}|g_{R}\rangle_{2}|g_{L}\rangle_{3}|g_{L}\rangle_{4}|g_{R}\rangle_{5}|g_{R}\rangle_{6}|g_{L}\rangle_{7}|g_{L}\rangle_{8}$,\\
$|g_{L}\rangle_{1}|g_{L}\rangle_{2}|g_{L}\rangle_{3}|g_{L}\rangle_{4}|g_{L}\rangle_{5}|g_{L}\rangle_{6}|g_{L}\rangle_{7}|g_{L}\rangle_{8}$,
$|g_{L}\rangle_{1}|g_{L}\rangle_{2}|g_{L}\rangle_{3}|g_{L}\rangle_{4}|g_{R}\rangle_{5}|g_{R}\rangle_{6}|g_{R}\rangle_{7}|g_{R}\rangle_{8}$,\\
$|g_{R}\rangle_{1}|g_{R}\rangle_{2}|g_{R}\rangle_{3}|g_{R}\rangle_{4}|g_{L}\rangle_{5}|g_{L}\rangle_{6}|g_{L}\rangle_{7}|g_{L}\rangle_{8}$, and
$|g_{R}\rangle_{1}|g_{R}\rangle_{2}|g_{R}\rangle_{3}|g_{R}\rangle_{4}|g_{R}\rangle_{5}|g_{R}\rangle_{6}|g_{R}\rangle_{7}|g_{R}\rangle_{8}$, the purification is successful. Similar with the case of $M=3$, the parties make all the photons pass through the HWPs. Then, they make each of the photons in the $c_{1}c_{2}c_{3}c_{4}$ and $d_{1}d_{2}d_{3}d_{4}$ modes pass through a PBS, and detect the photons in the output modes. They can finally obtain
\begin{eqnarray}
\rho_{out5}=F'|\Phi^{+}_{4}\rangle_{AB}\langle\Phi^{+}_{4}|+(1-F')|\Psi^{+}_{4}\rangle_{AB}\langle\Psi^{+}_{4}|
\end{eqnarray}
with $F'=\frac{F^{2}}{F^{2}+(1-F)^{2}}$, and the purification task is completed.

\subsection{The purification for the phase-flip error}
\begin{figure}[!h]
\begin{center}
\includegraphics[width=10cm,angle=0]{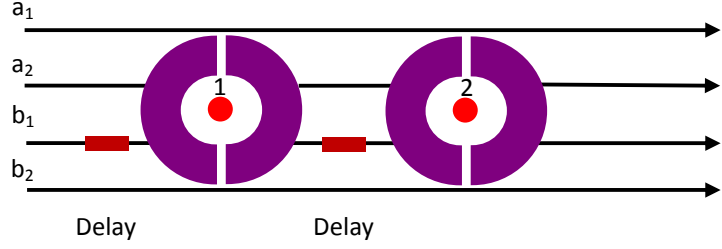}
\caption{The schematic drawing of the EPP for logic phase-flip error of the logic Bell state with $M=2$. The parties  need to prepare two three-level atoms with the form of $\frac{1}{\sqrt{2}}(|g_{L}\rangle+|g_{R}\rangle)$ trapped in four low-Q cavities, respectively. }
\end{center}
\end{figure}

Besides the bit-flip error, the phase-flip error is also unavoidable. We suppose a phase-flip error occurs with the probability of $1-F$, which will make $|\Phi^{+}\rangle_{AB}$ become
\begin{eqnarray}
|\Phi^{-}\rangle_{AB}=\frac{1}{\sqrt{2}}(|\phi^{+}\rangle_{A}|\phi^{+}\rangle_{B}-|\phi^{-}\rangle_{A}|\phi^{-}\rangle_{B}).
\end{eqnarray}
Under this case, the parties will obtain a mixed input state as
\begin{eqnarray}
\rho_{in2}=F|\Phi^{+}\rangle_{AB}\langle\Phi^{+}|+(1-F)|\Phi^{-}\rangle_{AB}\langle\Phi^{-}|.\label{mix2}
\end{eqnarray}

We will prove that for the logic Bell state, the logic phase-flip error can be completely corrected. The states $|\Phi^{\pm}\rangle_{AB}$ can be rewritten as
\begin{eqnarray}
&&|\Phi^{+}\rangle_{AB}=\frac{1}{\sqrt{2}}(|\phi^{+}\rangle_{A}|\phi^{+}\rangle_{B}+|\phi^{-}\rangle_{A}|\phi^{-}\rangle_{B})\nonumber\\
&=&\frac{1}{\sqrt{2}}[\frac{1}{\sqrt{2}}(|LL\rangle+|RR\rangle)_{a1a2}\frac{1}{\sqrt{2}}(|LL\rangle+|RR\rangle)_{b1b2}
+\frac{1}{\sqrt{2}}(|LL\rangle-|RR\rangle)_{a1a2}\frac{1}{\sqrt{2}}(|LL\rangle-|RR\rangle)_{b1b2}]\nonumber\\
&=&\frac{1}{\sqrt{2}}(|LLLL\rangle_{a1a2b1b2}+|RRRR\rangle_{a1a2b1b2}),\\
&&|\Phi^{-}\rangle_{AB}=\frac{1}{\sqrt{2}}(|\phi^{+}\rangle_{A}|\phi^{+}\rangle_{B}-|\phi^{-}\rangle_{A}|\phi^{-}\rangle_{B})\nonumber\\
&=&\frac{1}{\sqrt{2}}[\frac{1}{\sqrt{2}}(|LL\rangle+|RR\rangle)_{a1a2}\frac{1}{\sqrt{2}}(|LL\rangle+|RR\rangle)_{b1b2}
-\frac{1}{\sqrt{2}}(|LL\rangle-|RR\rangle)_{a1a2}\frac{1}{\sqrt{2}}(|LL\rangle-|RR\rangle)_{b1b2}]\nonumber\\
&=&\frac{1}{\sqrt{2}}(|LLRR\rangle_{a1a2b1b2}+|RRLL\rangle_{a1a2b1b2}).\label{phase}
\end{eqnarray}
 As shown in Fig. 5, the parties only need to prepare two three-level atoms with the form of $\frac{1}{\sqrt{2}}(|g_{L}\rangle+|g_{R}\rangle)$ trapped in two low-Q cavities, respectively. Then, they make the photons in the $a_{2}b_{1}$ modes pass through the two cavities and interact with the two atoms "1" and "2", successively. After both the two photons exit the cavities, the parties make a Hadamard operation on both the two atoms and then measure them in the basis of $\{|g_{L}\rangle, |g_{R}\rangle\}$. According to the input-output relations in Eq. (\ref{relation1}), if the measurement result of the two atoms is $|g_{L}\rangle$, the parties can ensure that the photon state is $|\Phi^{-}\rangle_{AB}$, which indicates a logic phase-flip error occurs. If the measurement result is $|g_{R}\rangle$, the photon state must be $|\Phi^{+}\rangle_{AB}$, that is, no logic phase-flip error occurs. According to Eq. (\ref{phase}), when the logic phase-flip error occurs, the parties can completely correct the error by make bit-flip operations on the photons in $b_{1}b_{2}$ modes.

We can also extend the protocol for the logic phase-flip error to the logic Bell states with each logic qubit being the arbitrary GHZ state. Under this case, if the phase-flip error occurs with the possibility of $(1-F)$, $|\Phi^{+}_{M}\rangle_{AB}$ will convert to $|\Phi^{-}_{M}\rangle_{AB}$ with the form of
$|\Phi^{-}_{M}\rangle_{AB}=\frac{1}{\sqrt{2}}(|GHZ^{+}_{M}\rangle_{A}|GHZ^{+}_{M}\rangle_{B}-|GHZ^{-}_{M}\rangle_{A}|GHZ^{-}_{M}\rangle_{B})$. Therefore, the parties will obtain a mixed state as
\begin{eqnarray}
\rho_{M1}=F|\Phi^{+}_{M}\rangle\langle\Phi^{+}_{M}|+(1-F)|\Phi^{-}_{M}\rangle\langle\Phi^{-}_{M}|.\label{mixM}
\end{eqnarray}

The discrimination process for $|\Phi^{+}_{M}\rangle_{AB}$ and $|\Phi^{-}_{M}\rangle_{AB}$ are quite similar with that for distinguishing $|\Phi^{+}\rangle_{AB}$ and $|\Phi^{-}\rangle_{AB}$. We can rewrite $|\Phi^{\pm}_{M}\rangle_{AB}$ as
\begin{eqnarray}
&&|\Phi^{+}_{M}\rangle_{AB}=\frac{1}{\sqrt{2}}(|GHZ^{+}_{M}\rangle_{A}|GHZ^{+}_{M}\rangle_{B}+|GHZ^{-}_{M}\rangle_{A}|GHZ^{-}_{M}\rangle_{B})\nonumber\\
&=&\frac{1}{\sqrt{2}}[\frac{1}{\sqrt{2}}(|LL\cdots L\rangle+|RR\cdots R\rangle)_{a1a2\cdots am}\frac{1}{\sqrt{2}}(|LL\cdots L\rangle+|RR\cdots R\rangle)_{b1b2\cdots bm}\nonumber\\
&+&\frac{1}{\sqrt{2}}(|LL\cdots L\rangle-|RR\cdots R\rangle)_{a1a2\cdots am}\frac{1}{\sqrt{2}}(|LL\cdots L\rangle-|RR\cdots R\rangle)_{b1b2\cdots bm}]\nonumber\\
&=&\frac{1}{\sqrt{2}}(|LL\cdots L\rangle_{a1a2\cdots amb1b2\cdots bm}+|RR\cdots R\rangle_{a1a2\cdots amb1b2\cdots bm}),\\
&&|\Phi^{-}_{M}\rangle_{AB}=\frac{1}{\sqrt{2}}(|GHZ^{+}_{M}\rangle_{A}|GHZ^{+}_{M}\rangle_{B}-|GHZ^{-}_{M}\rangle_{A}|GHZ^{-}_{M}\rangle_{B})\nonumber\\
&=&\frac{1}{\sqrt{2}}[\frac{1}{\sqrt{2}}(|LL\cdots L\rangle+|RR\cdots R\rangle)_{a1a2\cdots am}\frac{1}{\sqrt{2}}(|LL\cdots L\rangle+|RR\cdots R\rangle)_{b1b2\cdots bm}\nonumber\\
&-&\frac{1}{\sqrt{2}}(|LL\cdots L\rangle-|RR\cdots R\rangle)_{a1a2\cdots am}\frac{1}{\sqrt{2}}(|LL\cdots L\rangle-|RR\cdots R\rangle)_{b1b2\cdots bm}]\nonumber\\
&=&\frac{1}{\sqrt{2}}(|LL\cdots LRR\cdots R\rangle_{a1a2\cdots amb1b2\cdots bm}+|RR\cdots RLL\cdots L\rangle_{a1a2\cdots amb1b2\cdots bm}).
\end{eqnarray}
The parties prepare two three-level atoms with the form of $\frac{1}{\sqrt{2}}(|g_{L}\rangle+|g_{R}\rangle)$ trapped in two low-Q cavities, respectively, and make the photons in the $a_{m}b_{1}$ modes pass through the two cavities and interact with the two atoms "1" and "2", successively. After both the two photons exit the cavities, the parties make a Hadamard operation on both the two atoms and then measure them in the basis of $\{|g_{L}\rangle, |g_{R}\rangle\}$. If the measurement result of the two atoms is $|g_{L}\rangle$, the parties can ensure that the photon state is $|\Phi^{-}_{M}\rangle_{AB}$, that is, a logic phase-flip error occurs. If the measurement result is $|g_{R}\rangle$, the photon state must be $|\Phi^{+}_{M}\rangle_{AB}$, which indicates no logic phase-flip error occurs. Finally, under the case that the logic phase-flip error occurs, the parties can completely correct it by making bit-flip operations on the photons in $b_{1}b_{2}\cdots b_{m}$ modes.

\section{The purification of the physic-qubit error}
In the above section, we have successfully purified the bit-flip error and phase-flip error in the logic qubit. On the other hand, in the practical applications, the single physic qubit can also suffer from the bit-flip error or phase-flip error.

\begin{figure}[!h]
\begin{center}
\includegraphics[width=8cm,angle=0]{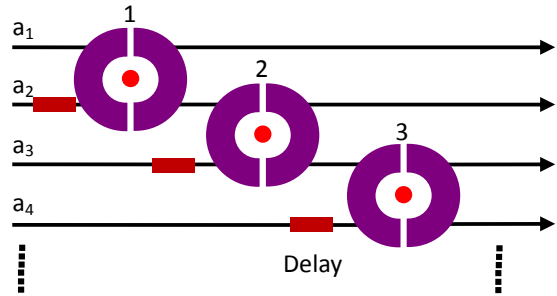}
\caption{The physic bit-error can be completely selected with the help of some three-level atoms with the form of $\frac{1}{\sqrt{2}}(|g_{L}\rangle+|g_{R}\rangle)$ trapped in the low-Q cavities. }
\end{center}
\end{figure}

For $|\Phi^{+}\rangle_{AB}$ in Eq. (\ref{initial}), we suppose a bit-flip error occurs in one of the physic qubits in the logic-qubit $A$. It makes one of the physic qubits in $|GHZ^{\pm}_{M}\rangle$ become $|L\rangle\leftrightarrow |R\rangle$. In this way, $|GHZ^{\pm}_{M}\rangle$ will change to $\frac{1}{\sqrt{2}}(|LL\cdots LRL\cdots L\rangle\pm|RR\cdots RLR \cdots R \rangle).$ As the error occurs locally, the parties can completely select the physic qubit with the bit-flip error by the local operations as shown in Fig. 6.

 They prepare some three-level atoms with the form of $\frac{1}{\sqrt{2}}(|g_{L}\rangle+|g_{R}\rangle)$ trapped in the low-Q cavities, respectively. They first make the photons in the spatial modes $a_{1}$ and $a_{2}$ pass through the cavity and interact with atom "1", successively. After the photon in $a_{2}$ exiting the cavity, they make the Hadamard operation on atom "1" and then measure the state of it.
If the measurement result is $|g_{L}\rangle$, the polarization state of the photons in $a_{1}$ and $a_{2}$ modes must be different. In this way, a bit-flip error occurs on the photon in $a_{1}$ or $a_{2}$ mode. In order to ensure which photon has the bit-flip error, they make the photons in $a_{2}$ and $a_{3}$ modes enter the cavity and interact with the atom "2", successively. After the photon-atom interaction, they make the Hadamard operation on atom "2" and then measure the state of it. If it is also in $|g_{L}\rangle$, it means the polarization state of the photons in $a_{2}$ and $a_{3}$ modes are different. Under this case the bit-flip error must occur on the photon in $a_{2}$ mode. If the measurement result of atom "2" is $|g_{R}\rangle$, it means the polarization state of the photons in $a_{2}$ and $a_{3}$ modes are the same. They can confirm the bit-flip error occurs on the photon in $a_{1}$ mode. On the other hand, if the measurement result of atom "1" is $|g_{R}\rangle$, it means the polarization state of the photons in $a_{1}$ and $a_{2}$ modes are the same, that is, no bit-flip error occurs on the photons of $a_{1}$ and $a_{2}$ modes. Next, they make the photons in $a_{2}$ and $a_{3}$ modes interact with atom "2", successively. After the interaction, if the measurement result of atom "2" is $|g_{L}\rangle$, the bit-flip error occurs on the photon in $a_{3}$ mode. If the measurement of atom "2" is $|g_{R}\rangle$, it means no bit-flip error occurs on the photons in $a_{1}a_{2}a_{3}$ modes. Next, the parties continue to make the photons in $a_{3}$ and $a_{4}$ modes interact with atom "3", successively, and so forth. Once the measurement result of atom "n" is $|g_{L}\rangle$,  the bit-flip error occurs on the photon in $a_{n+1}$ mode. In the whole process, the parties require two atoms at least corresponding to the error in $a_{1}$ or $a_{2}$ modes and $M-1$ atoms at most corresponding to the error in $a_{m}$ mode. After selecting the error photon, we can correct the error with a bit-flip operation. Similarly, if a bit-flip error occurs on the second logic
qubit $B$, they can also completely correct it with the same principle. As the parties only require to measure the atom state, the corrected photon state can be remained perfectly for other applications.

On the other hand, we consider a phase-flip error occurs in the logic-qubit $A$, which makes $|GHZ^{+}_{M}\rangle \leftrightarrow |GHZ^{-}_{M}\rangle$. Under this case, the $|\Phi^{+}_{M}\rangle_{AB}$ will change to $|\Phi^{+}_{M}\rangle_{AB}\rightarrow\frac{1}{\sqrt{2}}(|GHZ^{-}_{M}\rangle_{A}|GHZ^{+}_{M}\rangle_{B}+|GHZ^{+}_{M}\rangle_{A}|GHZ^{-}_{M}\rangle_{B})$. Obviously,the phase-flip error in logic-qubit $A$ equals to the bit-flip error in the logic entanglement shown in Eq. (\ref{initialN}). Therefore, the parites can complete the purification based on the EPP described in above.

\section{Discussion and conclusion}
In common physic-qubit entanglement, there are only two kinds of errors, say bit-flip error and phase-flip error. The traditional EPPs for the physic-qubit entanglement can directly purify the bit-flip error. For the phase-flip error, they need to convert it to the bit-flip error first, and purify it in the next step. For the logic-qubit entanglement, the error modes are more complicated. There are four different kinds of errors, say the bit-flip error and phase-flip error in the logic-qubit entanglement, and the bit-flip error and phase-flip error in the physic-qubit entanglement, respectively. In the paper, we put forward an effective EPP for dealing with the degraded logic Bell state with arbitrary $M$ based on the atom-photon interaction in the low-Q cavity. In the protocol, the phase-flip error in one physic qubit of a logic qubit can be transformed to the logic bit-flip error. The logic bit-flip error can be purified directly. The parties require two copies of the initial mixed photon states. For completing the purification task, they need to prepare $2M$ ($M$ is even) or $2(M+1)$ ($M$ is odd) three-level atoms in the form of $\frac{1}{\sqrt{2}}(|g_{L}\rangle+|g_{R}\rangle)$ trapped in the low-Q cavities, respectively. They make the photons enter the cavities and interact with the atoms successively. After the photon-atom interaction, the atomic states of all the atoms are measured in the basis of $\{|g_{L}\rangle, |g_{R}\rangle\}$. Then, they measure the states of the second copy of photon states. Based on the atomic and photonic measurement results, the parties can complete the purification task. The fidelity of the new mixed state ($F'$) is higher than that of the initial mixed state ($F$), under the case that $F>\frac{1}{2}$. We also prove that the logic phase-flip error and physic bit-flip error in one of the physic qubits of a logic qubit can be completely corrected with the help of some auxiliary three-level atoms in the low-Q cavities and the bit-flip operation. In this way, our protocol  can completely deal with all the four kinds of errors of the arbitrary logic Bell state. Moreover, as the parties only measure the atom states, all the distilled photon states can be well remained for other applications.

The key element of the EPP is the low-Q cavity. As shown in Sec. 2, in order to obtain the input-output relationship as Eq. (\ref{relation}), we must control the frequency of the input coherent state $\omega_{p}$ to meet $\omega_{p}=\omega_{c}-\frac{\kappa}{2}=\omega_{0}-\frac{\kappa}{2}$, which ensure $\theta=\pi$ and $\theta_{0}=\frac{\pi}{2}$.
Current research showed that the single-electron spin in a single quantum dot inside a micro-cavity and the
nitrogen-vacancy ($N-V$) defect center in diamond can induce the giant optical Faraday rotation \cite{experiment1,experiment2,experiment3}. In this way, our EPP can also be suitable for entangled electrons using a
quantum dot and microcavity coupled system. Recently, some attractive experiment results about the low-Q cavity have also been reported. For example, in 2005, Nu$\beta$mann \emph{et al.} showed that they can precisely control and adjust
the individual ultracold $^{85}Rb$ atoms coupled to a high-finesse optical cavity \cite{Rb1}. The states of $|F\rangle=2$. $m_{F}=\pm1$ of the $5S_{1/2}$ are chosen to be the two ground
states $|g_{L}\rangle$ and $|g_{R}\rangle$, respectively. The transition frequency between the
ground states and the excited state at $\lambda=780 nm$ is $\omega_{0}=\frac{2 \pi c}{\lambda} \approx 2.42 \times 10^{15} Hz$. The
cavity length, cavity rate and the finesse are $L=38.6 \mu m$, $K = 2\pi \times 53 MHz$ and $F=37000$,
respectively. In 2007, the group of Fortier experimentally realized the deterministic loading of single $^{87}Rb$ atoms into the cavity by incorporating a deterministic loaded atom conveyor \cite{Rb2}. In the same year, Colombe \emph{et al.} also reported their experiment on realizing the strong atom-field coupling for Bose-Einstein condensates (BEC) in a fiber-based
F-P cavity on a chip \cite{Rb3}. They showed that the $^{87}Rb$ BEC can be positioned deterministically anywhere within
the cavity and localized entirely within a single antinode of the standing-wave cavity field. Based on these experimental achievements, our EPP may be experimentally realized in the near future.

In conclusion, in practical applications, four kinds of errors may occur in the logic Bell state, that is, the bit-flip error and phase-flip error in the logic qubits, and the bit-flip error and phase-flip error in the physic qubits. In the paper, with the help of the photon-atom interaction in low-Q cavity and atomic state measurement, we put forward an effective EPP to deal with the four kinds of errors in the logic Bell state, where each logic qubit is arbitrary M-particle GHZ state. For the logic bit-flip error, the parties require two copies of initial mixed photon states and some auxiliary single three-level atoms trapped in the low-Q cavities. With the help of the photonic Faraday rotation effect and the atomic state and photon state measurement, they can finally distill new mixed photon state. Under the case that the fidelity ($F$) of the initial photon state meets $F>\frac{1}{2}$, the fidelity ($F'$) of the distilled new mixed states is higher than $F$. The phase-flip error in the physic qubit equals to the bit-flip error in the logic qubits, which can also be purified with above EPP. On the other hand, we prove that the logic phase-flip error and a bit-flip error in one physic qubit can be selected with the help of some auxiliary three-level atoms in the low-Q cavities. Then, the parties can completely correct both the two kinds of errors with the physic bit-flip operation. In our protocol, all the distilled new photon states can be well remained for other applications. According to the above features, our EPP may be useful in the future long-distance quantum communication based
on logic-qubit entanglement.

\section*{ACKNOWLEDGEMENTS}
This work was supported by the National Natural Science Foundation
of China under Grant  Nos. 11474168 and 61401222, the Natural Science Foundation of
Jiangsu under Grant No. BK20151502, the Qing Lan Project in Jiangsu Province, and a Project
Funded by the Priority Academic Program Development of Jiangsu
Higher Education Institutions.\\

\end{document}